\documentclass[12pt,preprint]{aastex}

\usepackage{natbib}

\shorttitle{Method for Extracting Light Echo Fluxes}
\shortauthors{Newman and Rest}

\begin{document}

\title{A Method for Extracting Light Echo Fluxes Using
the NN2 Difference Imaging Technique}

\author{A.~B.~Newman}
\affil{Department of Physics, Washington University}
\affil{1 Brookings Drive, St. Louis, MO 63130}
\email{abnewman@wustl.edu}

\and \author{A.~Rest\altaffilmark{1}}
\affil{Cerro Tololo Inter-American Observatory}
\affil{National Optical Astronomy Observatory}
\affil{950 North Cherry Avenue, Tucson, AZ 85719}
\email{arest@ctio.noao.edu}
\altaffiltext{1}{Goldberg Fellow.}

 
\begin{abstract}
Light echoes are interesting because of the wealth of information they
offer about their progenitors and the reflecting dust. Due to their
faint surface brightnesses, difference imaging is
necessary to separate most light echoes from the sky background. However, difference
images reveal only the relative fluxes between two
epochs. Obtaining absolute fluxes for individual epochs has traditionally
relied on a single template image that is free of light echoes.
Since such an observation is normally unavailable, a light echo-free template
must be constructed by a complicated and usually subjective process.
Here we present an application of the NN2 method of
Barris et al.~to extract the relative fluxes of light echoes
across a range of epochs directly from a series of difference images. This
method requires no privileged image and makes maximal use of the
observational data. Statistical methods to estimate the zero-flux
level and thus the absolute flux are also presented. The efficacy of
the technique is demonstrated by an application to the light echoes
around SN 1987A. The resulting images reveal new detail and faint
light echo structures. This method can be adapted and applied to other
extended variable light sources, such as stellar outflows and supernova remnants.
\end{abstract}

\keywords{methods: data analysis --- supernovae: general ---
techniques: image processing --- techniques: photometric}



\section{Introduction}
\label{sec:intro}
Light echoes arise when light from a stellar event is reflected toward
the observer by intervening dust. The arrival of the light is delayed
by the extra distance it traverses and so may be observed long after
the source event. Light echoes provide a powerful tool for studying
interesting stellar phenomena and sensitively probing dust structures.
For many dust geometries of interest, the light echoes
appear as rings or arclets which usually expand superluminally \citep{Couderc39}.
Light echoes from variable stars have been observed, including Nova
Persei 1901 \citep{Ritchey1901}, the source of the first discovered light echo,
and the mysterious eruptive variable V838 Monocerotis \citep[e.g.,][]{Munari02,Bond03}.
However, only supernovae (SNe) have sufficient luminosity to produce observable
scattering from interstellar dust \citep{Sugerman03}. Most light echo
discoveries, therefore, have been associated with SNe, 
including SN 1991T \citep{Schmidt94}, SN 1993J \citep{Sugerman02},
SN 1998bu \citep{Cappellaro01}, SN 2003gd \citep{Sugerman03:discov}, and
most famously SN 1987A, whose light echoes have been used to investigate in
detail the geometry, density, and composition of the circumstellar and interstellar
media \citep{Crotts91, Xu94, Xu95, Crotts95, Xu99, Sugerman05:II}.
Contributions from light echoes may
significantly affect the observed light curves of Type II SNe
\citep[e.g.,][]{Chugai92,DiCarlo02}. Polarization measurements of light echoes
can provide an independent distance measurement when scattering is nearly
$90^{\circ}$ \citep{Sparks94, Sparks96}. \citet{Rest05:le} have analyzed the
spectra of light echoes from ancient SNe to determine the types of their
progenitors. Although the wealth of information they contain has inspired
searches \citep[e.g.,][]{Boffi99}, most light echoes are difficult to image
due to their low surface brightnesses.

\begin{figure}
\begin{center}
\includegraphics[scale=0.75]{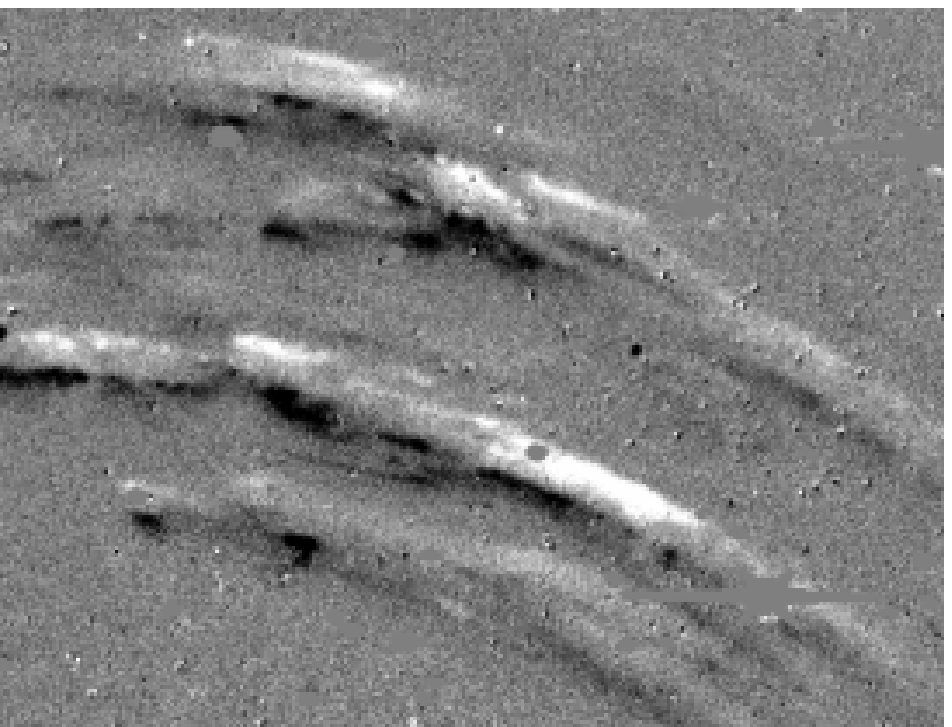}
\end{center}
\caption{Sample difference image showing 2005 December 31 (image epoch)
minus 2005 January 7 (template epoch). White pixels are positive and thus
enriched in the image epoch, while black pixels are negative
and enriched in the template epoch. Note that this image contains light
echo fluxes from both epochs which frequently overlap and exhibit
partial cancellation. The FOV is $3.1^{\prime} \times 2.3^{\prime}$.
\label{fig:diffim_example}}
\end{figure}

Sophisticated difference imaging techniques
\citep[e.g.,][]{Tomaney96,Alard98,Alard00,Sugerman05:II,Rest05:le} allow light echoes
to be cleanly separated from the sky background, but difference imaging alone
can reveal only the relative flux between two epochs. Each difference image
shows the flux difference between an image epoch and a template epoch.
If a light echo-free image is available, it can serve as a template
that can be subtracted from every epoch to isolate the light echoes.
Unfortunately, such an observation usually does not exist. 
Previous studies \citep[e.g.,][]{Xu95, Sugerman05:II} have therefore focused on
constructing a light echo-free image. The construction involves
a rather complicated iterative process, requires some subjective decisions,
and tends to leak faint light echoes into the echo-free template.
Another possibility is to create difference images from epochs that both
contain light echo flux. In this case,
the sky background is eliminated in the difference image, but both positive and negative
light echo fluxes remain. Figure~\ref{fig:diffim_example}
shows a sample difference image in which these positive and negative
fluxes are represented by white and black pixels, respectively.
Separating these frequently overlapping signals is a serious challenge.

In this paper, we present an application of the NN2 algorithm of
\citet{Barris05} to calculate relative light echo fluxes across a
range of epochs directly from a series of difference images.
This technique requires no light echo-free template
and treats all observations equally. It thus
makes maximal use of all the observational data.
We also present a statistical method for estimating the zero-flux level
of each pixel from the relative light curve, thus determining the absolute flux.
The entire procedure is tested on the light echoes around SN 1987A,
and the resulting images reveal intricate detail and faint structures.

\section{Observations and Reduction}
\label{sec:observations}
To test the ability of our technique to accurately recover light echo
fluxes, the light echoes around SN 1987A were chosen for
analysis. Observations were performed by the SuperMACHO survey on the Victor
M.~Blanco 4m Telescope at Cerro Tololo Inter-American Observatory
(CTIO). Imaging of the Large Magellanic Cloud (LMC) was performed
every other night in dark time during the months of October, November,
December, and January, with occasional images from September and
February, during five seasons from 2001 through 2005.  These are the
months in which the LMC is most visible from Cerro Tololo. We use the
$8K \times 8K$ Mosaic II CCD imager with a field of view (FOV) of 0.33 deg${}^2$,
along with a custom broadband \emph{VR} filter from 500nm to 750nm. For
this study, 52 epochs were chosen between 2001 November and 2005 December.
These were selected based on transparency, seeing conditions, 
eccentricity, and their temporal spacing. Images from these $N = 52$ epochs were
first reduced by the SuperMACHO pipeline. The reduction
includes bias correction, cross-talk elimination, flat-fielding, image
deprojection, and DoPHOT photometry \citep{Schechter93} to compute
the zero-point magnitude.

To isolate transient and variable objects such as light echoes, the SuperMACHO
pipeline includes difference image analysis. Point spread function (PSF) matched
subtraction is performed by the HOTPANTS software \citep{Becker04}. Since
the PSF can vary over the FOV due to optical distortions and
imperfect focus, a spatially varying kernel is employed \citep{Alard98,Alard00}.
For more information about the observation program, reduction, and difference
imaging, see \citet{Rest05:discrim}.

\section{The NN2 Difference Imaging Technique}
\label{sec:analysis}
The principal feature of the NN2 method is that it
considers difference images created from all $N(N-1)/2$ possible pairs of
observations at $N$ epochs. For this study, the SuperMACHO pipeline was used to produce the 1326
difference images that served as the input to the NN2 algorithm.
From the beginning, therefore,
we consider images that ideally contain only light echo signal, and the
goal is to seperate the positive and negative light echo fluxes contained in
these images.
The NN2 algorithm is a single-pixel approach. For each pixel, a fit in
$N - 1$ free parameters is performed to determine the relative fluxes
across the $N$ input epochs that best reproduce the flux differences found
in the full set of difference images. We thus isolate the light echo signal in each
epoch, without the need for an echo-free template.
After obtaining the relative fluxes, a separate
method is used to establish the absolute flux level for each pixel.
Finally, spatial and temporal binning are performed to increase image
quality and depth.

\subsection{The NN2 Algorithm}
\label{sec:nn2}
The NN2 algorithm was created by \citet{Barris05} to construct optimal
light curves for point-source variable objects from a series of
observations. To apply this technique to light echoes, we first
consider a single pixel.  Let $A$ be the antisymmetric matrix for
which $A_{ij}$ is flux in the difference image between epochs $j$ and
$i$. This flux is normalized to zero-point magnitude 30, based on the 
zero-point magnitudes of the images at epochs $j$ and $i$.
The matrix $A$, along with an error matrix $E$ whose entries contain
the uncertainties in the corresponding entries of $A$, represents the
input to the NN2 algorithm from the difference imaging pipeline. We
fit $A$ to the form $A_{ij} = V_j - V_i$ by minimizing
\begin{equation}
\label{tomin}
\chi^2 = \sum_{i < j} \frac{(-A_{ij} + V_j - V_i)^2}{E_{ij}^2} + \frac{\sum V_i}{\langle{E}\rangle^2},
\end{equation}
where $\langle{E}\rangle$ represents the average error, given by
\begin{equation}
\frac{1}{\langle{E}\rangle^2} = \frac{2}{N(N-1)} \sum_{i < j} \frac{1}{E_{ij}^2}.
\end{equation}
The vector $V$ then contains the relative fluxes across the $N$
epochs.  The second term in Equation~(\ref{tomin}) is necessary to
obtain a unique solution for $V$, which is otherwise insensitive to
the addition of a constant vector. This term implies $\sum V_i = 0$
and requires us to determine the zero-flux level separately for each
pixel.  As derived in \citet{Barris05}, $\chi^2$ is minimized by
solving a system of $N$ linear equations using matrix inversion.

There are two sources of uncertainty in $V$. The ``internal error''
is due to the fact that $A$ cannot in general be perfectly represented
as $A_{ij} = V_{j} - V_{i}$ for any $V$. An estimate for this error arises
naturally in the calculation of $V$, as described in \cite{Barris05}.
In addition, there are uncertainties in $V$
due to errors $E$ in the difference images, which we call the ``external
error.'' To estimate the external error in $V$, we assume that the uncertainty
$E_{ij}$ in the difference image arises from uncertainties $\sigma_i$ and
$\sigma_j$ in the original images at epochs $i$ and $j$, i.e., that
there exists a vector $\sigma$ such that
\begin{equation}
E_{ij}^2 = \sigma_i^2 + \sigma_j^2.
\end{equation}
To calculate the best-fit $\sigma$, we minimize
\begin{equation}
\label{chiemin}
\chi_e^2 = \sum_{i,j; i < j} \left(-1 + \frac{\sigma_i^2}{E_{ij}^2} + \frac{\sigma_j^2}{E_{ij}^2}\right)^2.
\end{equation}
The minimization condition
\begin{equation}
\frac{\partial \chi_e^2}{\partial \sigma_k^2} = 0
\end{equation}
results in a system of $N$ equations, which can be written as
\begin{equation}
\sum_{i; i \neq k} \frac{1}{E_{ik}^2} = \sum_{i} D_{ik} \sigma_i^2,
\end{equation}
where
\begin{eqnarray}
\displaystyle
D_{ik} &=& \frac{1}{E_{ik}^4}, \quad i \neq k \nonumber \\
D_{kk} &=& \sum_{i; i \neq k} \frac{1}{E_{ik}^4},
\end{eqnarray}
and solved by inverting $D$.

In rare cases, this method yields negative solutions for $\sigma_i^2$,
indicating that the original assumption that $E_{ij}^2$ can be
well-represented by $\sigma_i^2 + \sigma_j^2$ is not accurate. In
practice, this occurs in of order 1 in $10^5$ pixels, which are
therefore simply masked. The $\chi_e^2$ in equation~(\ref{chiemin}) is that
adopted in the NN2 code of M.~W.~Wood-Vasey, J.~Tonry, and
M.~C.~Novicki.\footnote{\texttt{See \url{http://ctiokw.ctio.noao.edu:8080/Plone/essence/Software/NN2}.}}
It differs slightly from the $\chi_e^2$ given in \cite{Barris05}.
In practice, the errors obtained are
similar when both methods are available, but minimizing the
$\chi_e^2$ of \citet{Barris05} fails far more often (i.e., yields negative
values for $\sigma_i^2$).

To compute the total uncertainty, the contributions of both the internal and
external errors must be considered. To combine these errors, we
use the weighted method presented in \citet{Barris05}, in which the 
internal error is weighted by the normalized $\chi^2$ of equation~(\ref{tomin}).

For the heart of the NN2 calculation, the \texttt{antivec} function,
we adapted the code of W.~M.~Wood-Vasey, J.~Tonry, and M.~C.~Novicki.  A region
of specified size is read into memory from each of the $N(N-1)/2$
difference images.  For each pixel, masked data are discarded as
follows. The number of images associated with each epoch (either as
the image or template date) in which the pixel is masked is
calculated. All images associated with the epoch having the highest
count are discarded, and the process is repeated until no masked data
remains. We consider input pixels as masked if they are marked by the
SuperMACHO pipeline as saturated or bad pixels.
The NN2 algorithm is run with the
remaining data, and a crude zero-flux level is set by shifting the
relative fluxes so that the minimum flux for each pixel is zero. Image files containing
the flux, internal error, external error, and total error are
created. In addition, mask files are created for each epoch showing
which pixels were discarded. Pixels that encountered an error in
calculating the external error are also flagged. From the mask files, an
image is created, indicating the total number of epochs included in
the calculation for each pixel. Pixels calculated with fewer than 30
epochs were masked in our study of SN 1987A.

\subsection{Zero-Flux--Level Determination}
\label{sec:zeroflux}
\begin{figure}
\begin{center}
\begin{tabular}{cc}
\includegraphics[scale=0.65]{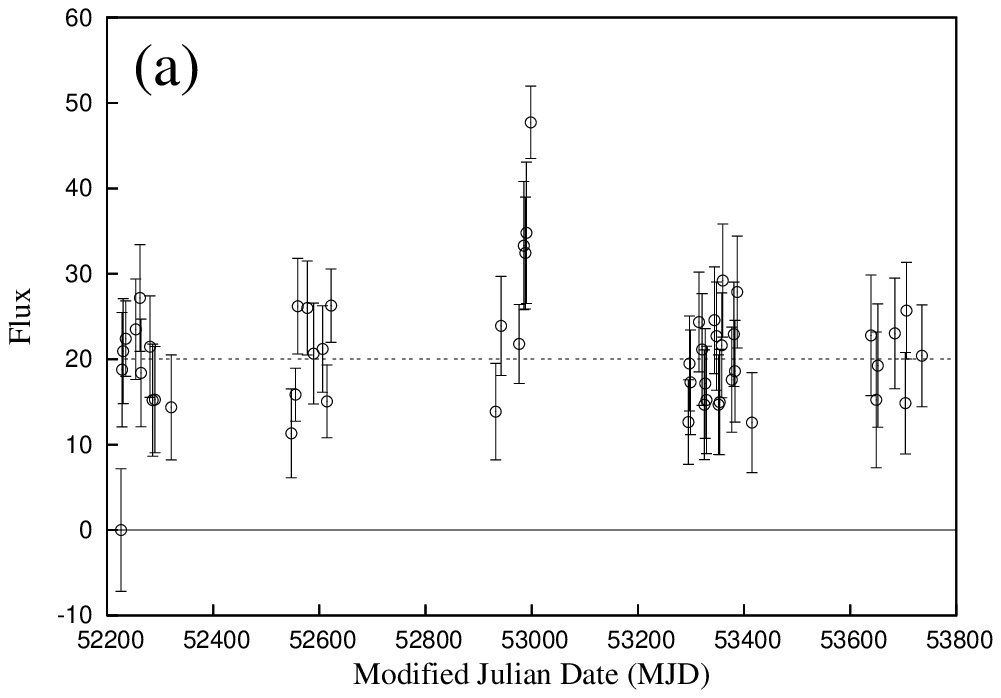} \\
\includegraphics[scale=0.65]{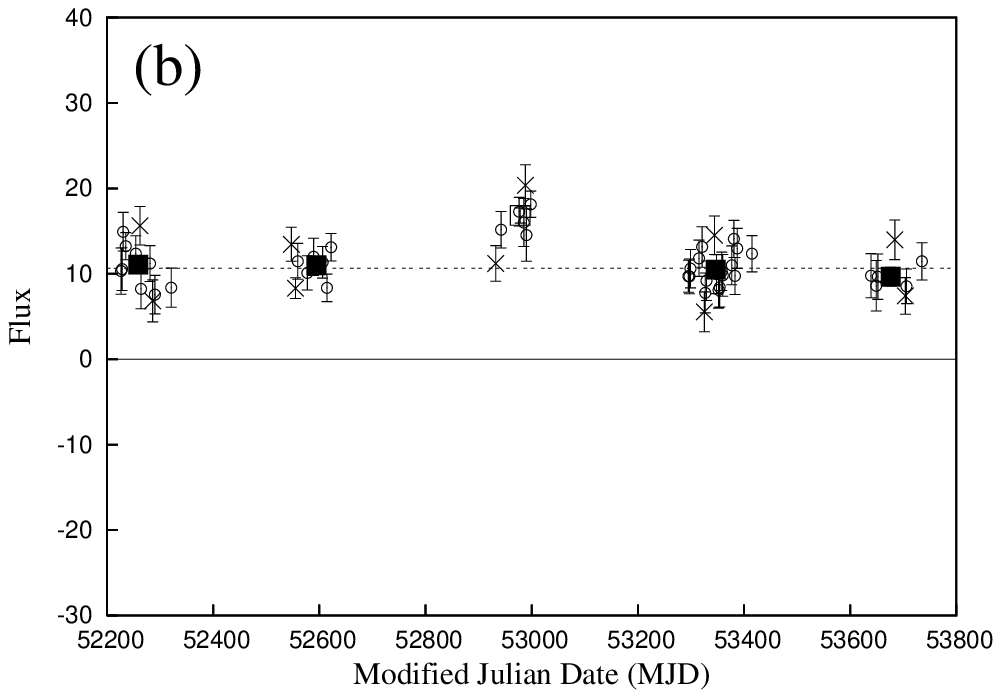} \\
\includegraphics[scale=0.65]{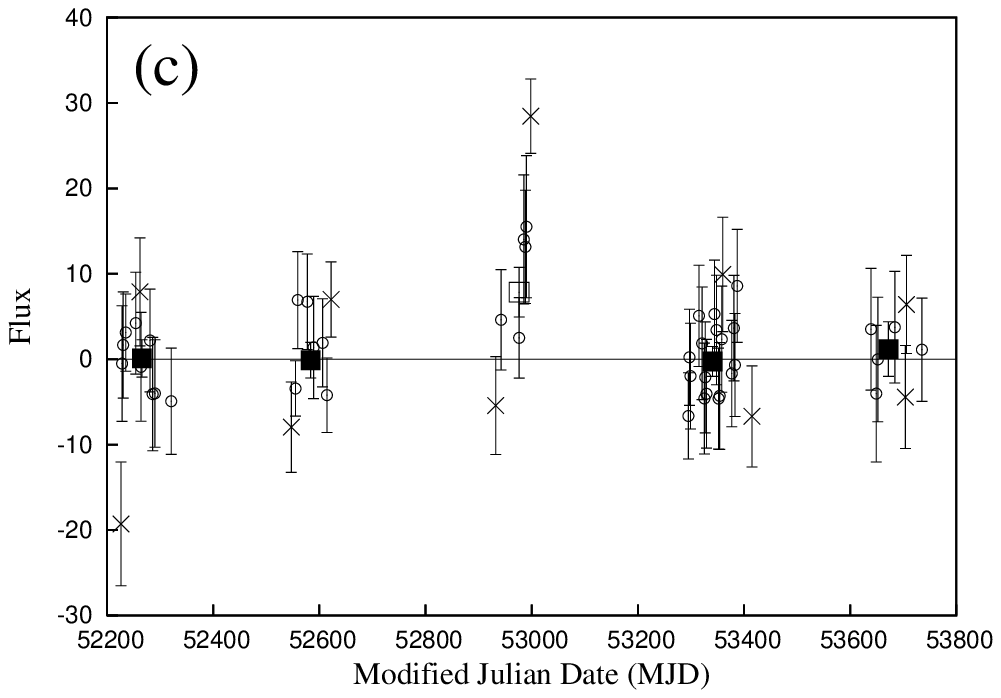}
\end{tabular}
\end{center}
\caption{Light curves for a pixel containing a faint light echo in season 3.
(\emph{a}) Relative fluxes obtained from the NN2 algorithm. (\emph{b}) Smoothed fluxes
obtained by spatially binning pixels in a $3 \times 3$ box. The crosses
denote the highest and lowest flux values from each season,
which are excluded in calculating the season means, denoted by squares.
A light echo is identified in season 3 by the sigma cut, and the remaining
four seasons are deemed to represent the zero-flux level. The dotted
line shows their mean.
(\emph{c}) Zero-flux--corrected light curve. The mean of the unbinned fluxes in the
zero-flux seasons is set as the zero-flux
level, the dotted line in (\emph{a}). This is subtracted from each epoch
to produce the corrected light curve here. Squares and crosses have the
same meaning as above.
Error bars show the 1 $\sigma$ uncertainties.
\label{fig:lightcurve_weakecho}}
\end{figure}

Since the NN2 algorithm provides only relative fluxes, the light curves
obtained for each pixel contain arbitrary offsets with respect to both 
the true zero-flux level and the neighboring pixels. It is therefore
necessary to estimate the true zero-flux level for each pixel.
Figure~\ref{fig:lightcurve_weakecho}\emph{a} shows the relative
light curve of an example pixel as determined by the NN2 algorithm.  As
a first crude estimate, the zero-flux level is taken to be the minimum
flux over all epochs. This value is subtracted from each epoch to
produce Figure~\ref{fig:lightcurve_weakecho}\emph{a}. Since our observations were taken in annual
spans of approximately 3 months for 5 years, we group the data
into five seasons. The example in Figure~\ref{fig:lightcurve_weakecho}
shows a very faint light echo in the third season. In order to
estimate the zero-flux level, we must identify the time periods (in
our case, seasons) that contain no light echo flux. For faint light
echoes, as in this example, this is very difficult, because the
individual pixel values are sometimes barely 1 $\sigma$ above the sky
background. To assist in
identifying the zero-flux epochs, we use the following properties of
light echoes:
\begin {itemize}
\item Light echoes are spatially extended, so nearby pixels are very likely
to share the same epochs with zero light echo flux. Thus, we can
spatially bin the fluxes by replacing each pixel with the weighted
mean of its unmasked neighbors within a $3 \times 3$ box, excluding
the highest and lowest fluxes. (Throughout this paper, extreme fluxes
are excluded from means only when at least six data points are available.)
The resulting spatially-smoothed light curve is shown in
Figure~\ref{fig:lightcurve_weakecho}\emph{b}. 
\item In most cases, light echoes do not change significantly over a time scale
of months. Thus we can temporally bin the fluxes as well. In our case,
the five seasons form the natural binning periods. Within each season,
the epochs of highest and lowest flux are excluded. These are denoted by
crosses in Figure~\ref{fig:lightcurve_weakecho}\emph{b}.
Using the remaining data, the weighted mean flux of each season is calculated.
These are shown as squares in Figure~\ref{fig:lightcurve_weakecho}.
In this example,
the mean of the spatially-binned third season is $6.4$ $\sigma$ above the mean
of the other four seasons. Without spatial binning, this is reduced to
$2.6$ $\sigma$.
\end{itemize}

Using the spatially- and temporally-binned light curve, we use the
following algorithm to determine the seasons that contain zero light
echo flux:
\begin{itemize}
\item Seasons whose mean is more than 3 $\sigma$ above the minimum mean among
all seasons are considered to contain light echoes and are thus
excluded. In practice, most seasons that will be cut are excluded by
this sigma cut.
\item Seasons that exhibit a strong gradient are excluded. The presence
of a gradient is detected by measuring how well the best-fit line fits
the data compared to a flat line at the mean. In particular, the ratio
of the $\chi^2$ statistic for the mean to
the $\chi^2$ statistic for the best-fit line, as
determined by the least-squares method,
is calculated. Seasons for which this ratio exceeds 10 are cut.
\item Seasons with a normalized $\chi^2$ greater than 3.5 are excluded.
\end{itemize}
The remaining seasons are deemed to contain zero light echo flux. In
the example of Figure~\ref{fig:lightcurve_weakecho}, these are seasons
1, 2, 4, and 5. (Season 3 is excluded by the sigma cut.) Since at least one
season must be used to calculate the zero-flux level, none
of the above steps is allowed
to exclude all the remaining seasons. If one initially does so, the
parameter being measured is examined, and the season that deviates
least from a perfect zero-flux season is retained. The critical values of
the parameters in the second and third steps were chosen by sampling pixels
from regions containing passing light echoes and choosing values to best
reproduce the zero-flux level seen in the light curve. They can be adjusted
for different data sets; however, since the second and third steps are included
as safeguards after the dominant sigma cut, the exact critical values are not
likely to significantly affect the output.

Having determined the zero-flux seasons, we use this information to
correct the original, unbinned light curve, shown in
Figure~\ref{fig:lightcurve_weakecho}\emph{a}. For each zero-flux season,
the highest and lowest fluxes are cut to reduce the influence of outlier data. 
(We deemed this cut sufficient for our data. Those planning
future applications may want to consider an iterative sigma clip instead.)
The weighted mean of all zero-flux seasons is taken to be the
zero-flux level, shown by the dotted line in Figure~
\ref{fig:lightcurve_weakecho}\emph{a}. This value is subtracted from the flux at every
epoch, and its uncertainty is combined quadratically with errors in
the fluxes, to produce the zero-flux corrected light curve, as shown in
Figure~\ref{fig:lightcurve_weakecho}\emph{c}.

\begin{figure}
\centering
\begin{tabular}{c}
\includegraphics[scale=0.75]{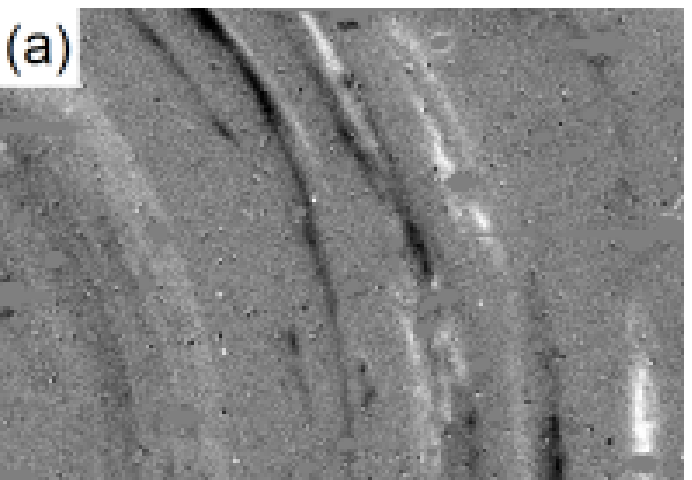} \\
\includegraphics[scale=0.75]{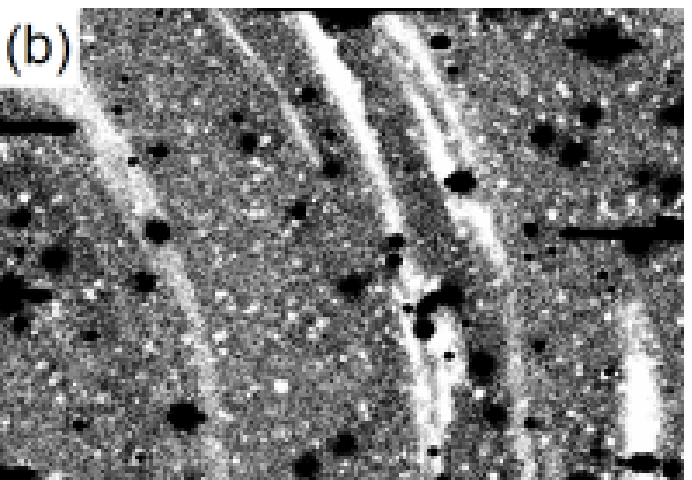} \\
\includegraphics[scale=0.75]{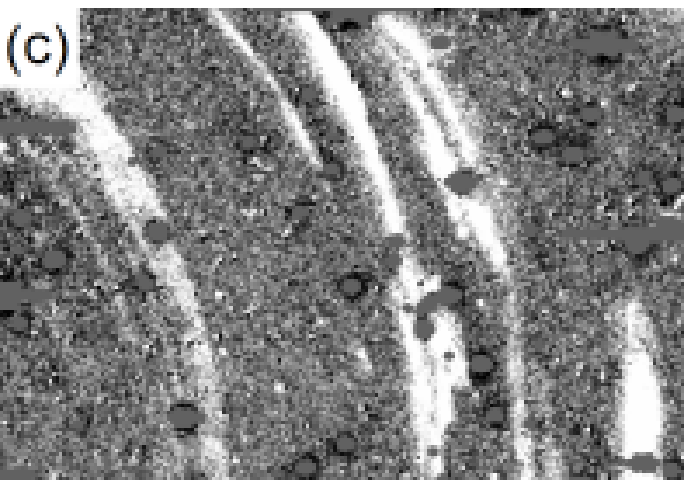}
\end{tabular}
\caption{(\emph{a}) Difference image with image and template epochs
of 2004 December 13 and 2001 December 11, respectively.
(\emph{b}) NN2 output for the image epoch of the difference image. For each pixel,
the relative light curve obtained from the NN2 algorithm is shifted
so that the minimum flux over all epochs is zero. Note that light echo
flux from the template epoch in (\emph{a}) has been eliminated.
(\emph{c}) Zero-flux--corrected image for the same epoch. Each pixel level has been
shifted to the proper flux zero point.
The FOV is $3.0^{\prime} \times 2.0^{\prime}$.
\label{fig:zerofluxlevel}}
\end{figure}

The results of this method, applied not only to a single pixel but to
an entire region, are shown in Figure~\ref{fig:zerofluxlevel}. Panel \emph{a}
shows a difference image, panel \emph{b} shows the NN2
output for the image epoch of the difference image, and panel \emph{c}
shows the zero-flux--corrected image for the same epoch.  Note
how much more clearly the light echoes appear above the sky background
in panel \emph{c}, especially the fainter features.

\begin{figure}
\centering
\begin{tabular}{cc}
\includegraphics[scale=0.75]{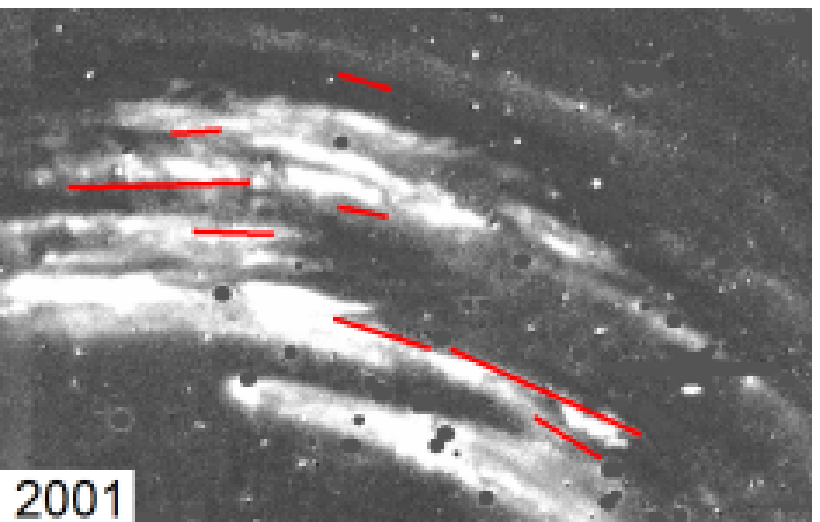} &
\includegraphics[scale=0.75]{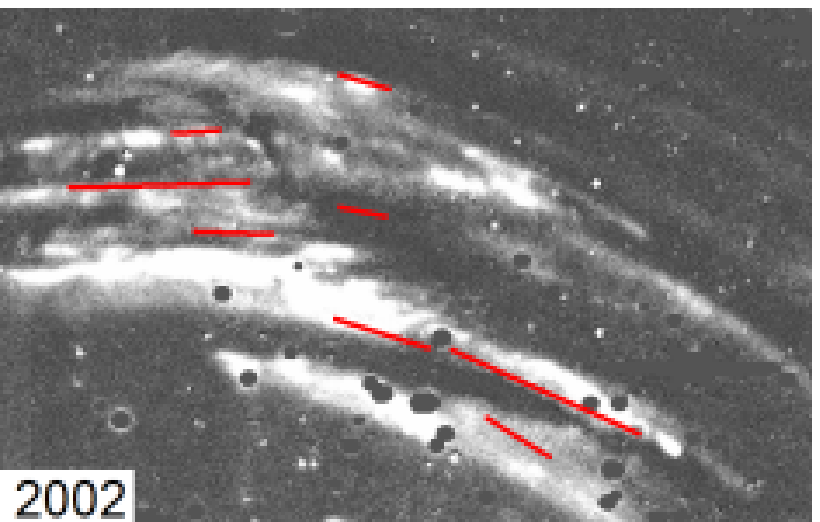} \\
\includegraphics[scale=0.75]{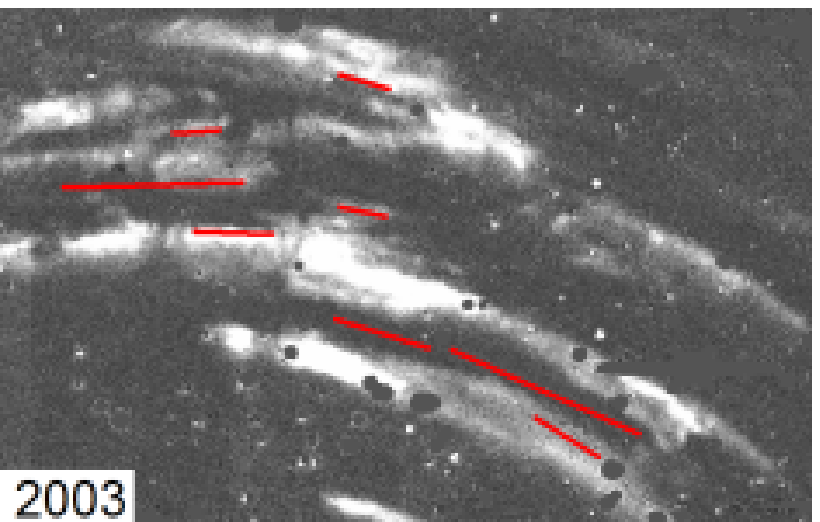} &
\includegraphics[scale=0.75]{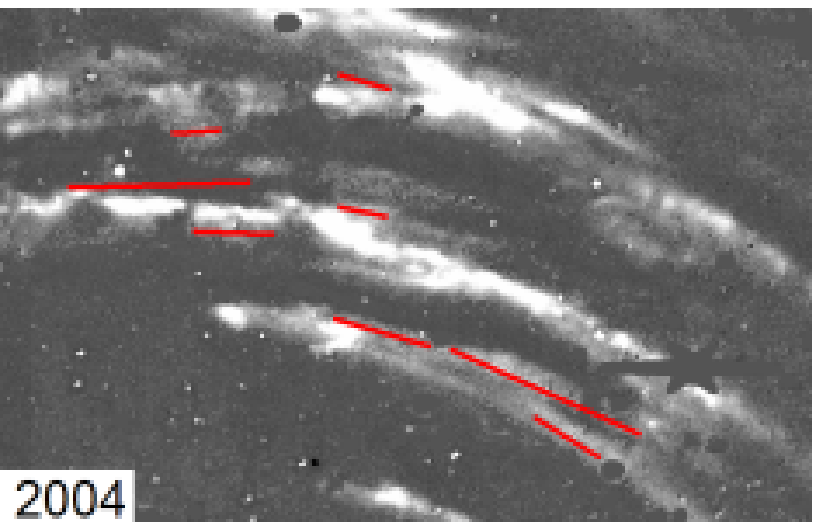} \\
\includegraphics[scale=0.75]{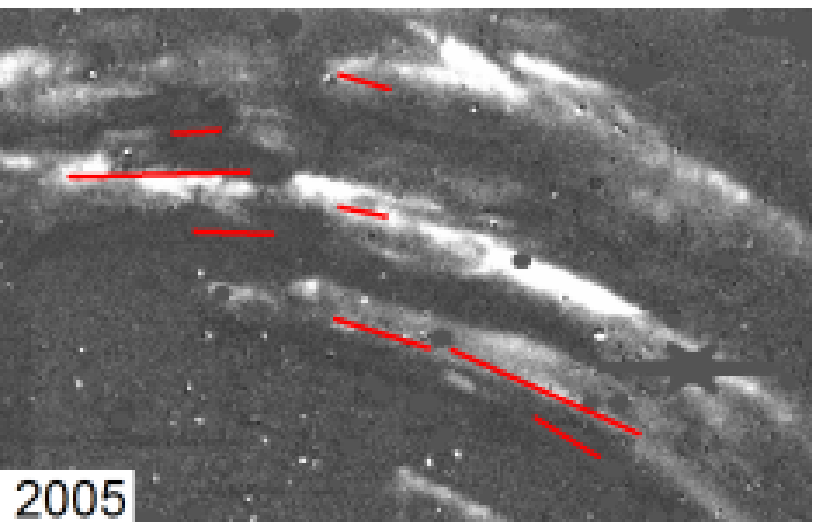} &
\includegraphics[scale=0.75]{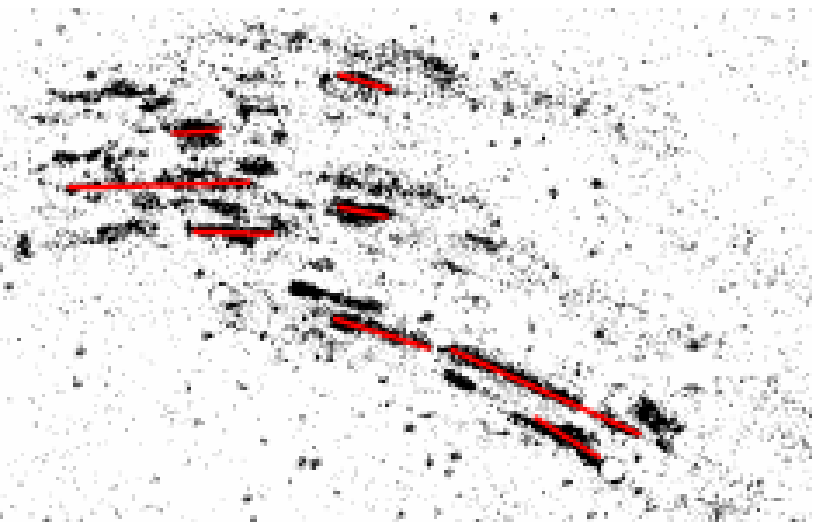}
\end{tabular}
\caption{Mean-combined images for the 2001--2005 seasons
showing a region with closely spaced light echo arcs. The final frame
shows those pixels for which only one zero-flux season is identified. The close
spacing of the light echo arcs in this region causes an unusually high number
of pixels to contain light echoes in all but one season. 
For these pixels, the flux values obtained here may be only
lower limits on the true flux. The red lines are constant
in each image and highlight areas dense in pixels with one zero-flux season.
Note that every such pixel necessarily has a calculated flux of zero in
one season, even if light echoes were physically present throughout the
observation period. By following the motion of the light echoes, however,
we can generally exclude this possibility. The FOV is $3.5^{\prime} \times 2.2^{\prime}$.
\label{fig:onebin}}
\end{figure}

The procedure described here is dependent on the availability of true
zero-flux epochs in the neighborhood of every pixel. For some pixels,
only one season is identified as containing zero light echo flux. For
them, it is possible that all seasons actually contain light echoes,
and that the identified season really contains the faintest light
echo region, rather than zero flux. In this case, the calculated
zero-flux level is an overestimate of the true value, and the fluxes
obtained are lower limits on the true fluxes. In our example of SN
1987A, the rings move quickly enough and are thin enough that few
pixels should contain light echoes for 5 consecutive
years. However, there are some regions that contain several closely
spaced light echo arclets. An extreme case is shown in
Figure~\ref{fig:onebin}, in which the final frame indicates those
pixels with only one zero-flux season identified. The red lines are constant
and indicate regions that are dense in such pixels. It is possible that
some of these pixels are illuminated by light echoes throughout the
observation period. The season during which the faintest echoes pass
is used to calculate the zero-flux level, and consequently the light
echoes disappear from the zero-flux--corrected images for that
season. This effect must be accounted for, especially if the technique
presented here is applied to light echoes with more complicated
structure.

\begin{figure}
\centering
\includegraphics[scale=0.7]{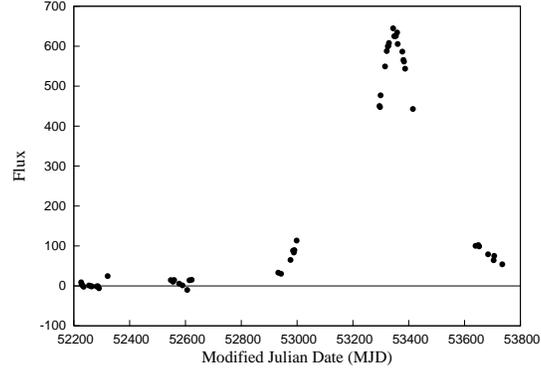}
\caption{Zero-flux--corrected light curve for a pixel containing a bright light echo.
Seasons 2 through 5 are excluded by the sigma cut, and the zero-flux level is determined
by season 1 alone. Nevertheless, given the smooth nature of the curve and the flatness of the
season 1 data, it is very likely that season 1 represents the true zero-flux level.
Uncertainties are comparable to the size of the points.
\label{fig:lightcurve_strongecho}}
\end{figure}

In our images of SN 1987A, 7\% of all usable pixels have only one
zero-flux season identified.  In all cases we have examined, this
season does seem to represent the true zero-flux level.  Sometimes
this can be verified from the
light curve. Figure~\ref{fig:lightcurve_strongecho} shows the zero-flux--corrected
light curve for a pixel containing a very bright light echo.
Despite being the only zero-flux season identified, the first season
does appear to contain zero light echo
flux. Pixels with only one zero-flux season can also be inspected by
following the motion of light echo arcs. If the zero-flux season falls
in a gap between arcs
that appears in images from the other seasons, we can be
confident in its veracity. If, on the other hand, a conspicuous hole
were to appear in the zero-flux season, altering the shape of the
light echo, the calculated fluxes would be interpreted as lower limits.

\subsection{Stacking and Smoothing}
\label{sec:stacksmooth}
\begin{figure}
\centering
\begin{tabular}{cc}
\includegraphics[scale=0.75]{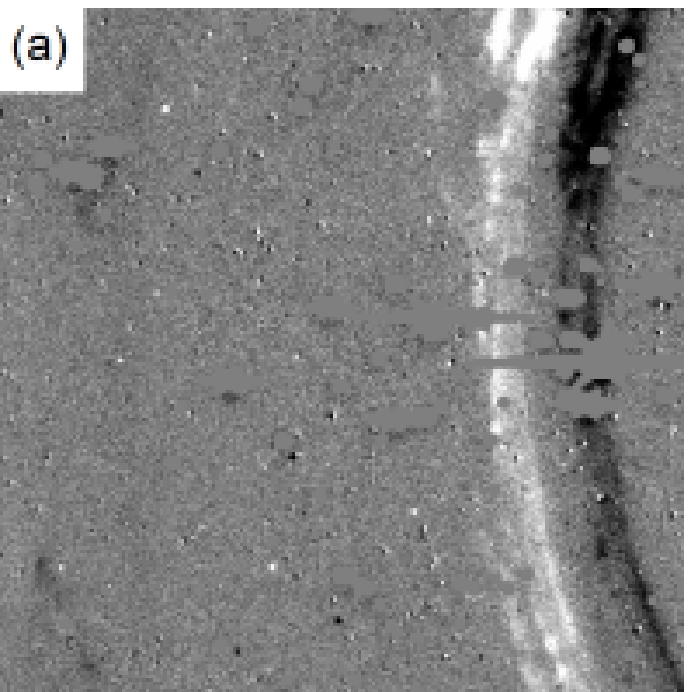} &
\includegraphics[scale=0.75]{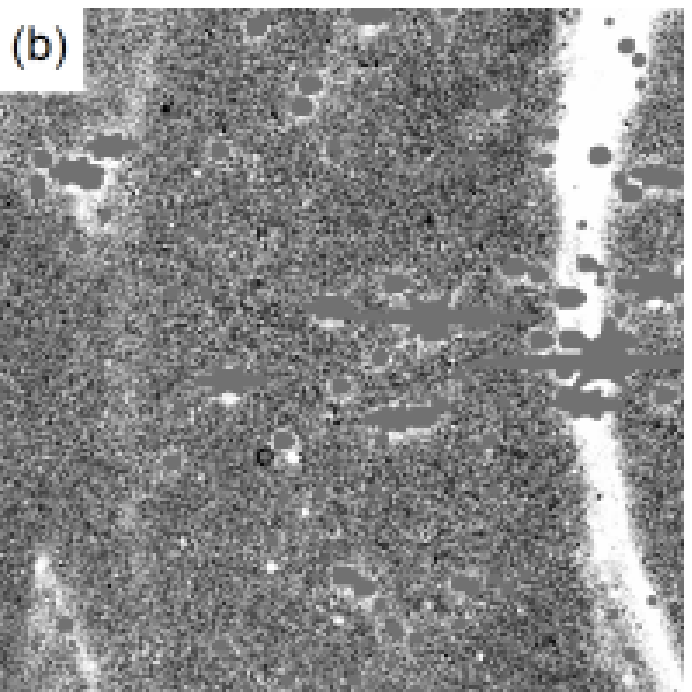} \\
\includegraphics[scale=0.75]{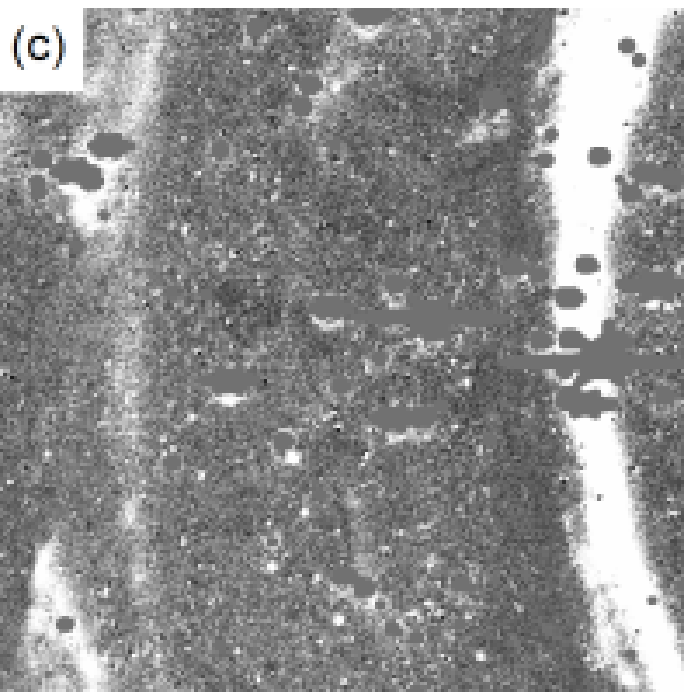} &
\includegraphics[scale=0.75]{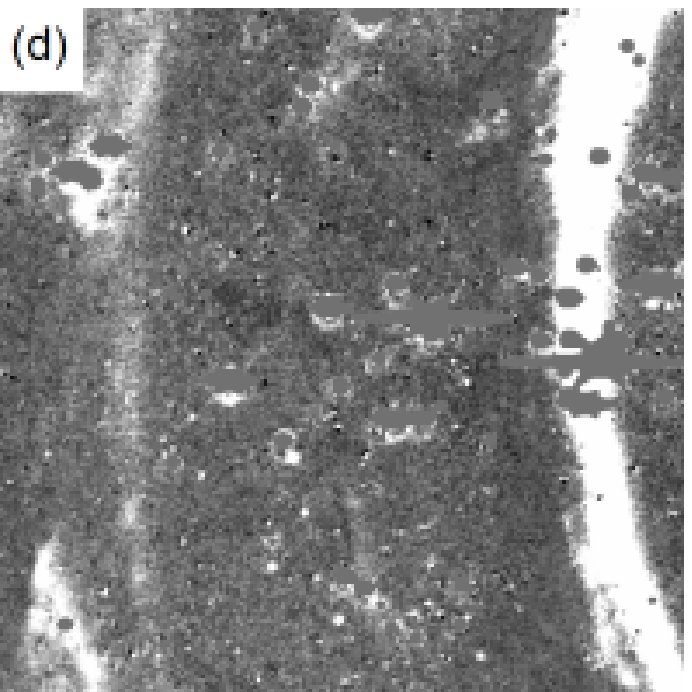}
\end{tabular}
\caption{Stages of light echo extraction
showing a region containing a weak light
echo arc in the left part of the frame. The FOV is $2.9^{\prime} \times 2.9^{\prime}$.
(\emph{a}) Difference image showing 2004 December 15 minus 2001 December 21.
(\emph{b}) Zero-flux corrected image for 2001 December 21. This is the template epoch of the difference image, in which it appears black.
(\emph{c}) Mean-combined image for the 2001 season.
(\emph{d}) Smoothed mean-combined image for the same season.
\label{fig:weakringseries}}
\end{figure}

To assist in separating very faint light echo pixels from the sky background, the
NN2 images from each season were mean-combined (stacked), excluding masked
epochs as well as those containing the highest and lowest fluxes,
resulting in five stacked images. Corresponding error and mask images
were also created.  Since the flux can change by a factor of 2 or more
within a single season in a region with a moving light echo, this
binning reduces the detail contained in the flux information. However, it
does increase the signal-to-noise ratio and improve the visibility
of faint rings for detection. The number of bad pixels with extreme
values is also reduced. These can arise, for example, from difference
imaging artifacts in the original difference images.  To further improve
visibility, the five stacked images were smoothed using the same spatial
binning technique described in section
\ref{sec:zeroflux}. The results are shown in
Figure~\ref{fig:weakringseries}. Panels \emph{a} and \emph{b} show a sample difference
image and the zero-flux--corrected output for its template epoch,
respectively. Panel \emph{c} shows the season mean-combined
image, formed by stacking many images like the one in \emph{b},
and \emph{d} shows the same image after spatial smoothing.
The stacking and smoothing procedure
typically increased the signal-to-noise ratio by a factor of 10 in our
images of SN 1987A. This could be anticipated, since stacking should
reduce the noise by a factor of approximately $(N/n)^{1/2}$, where $n$
is the number of seasons, and smoothing should further reduce the
noise by a factor of $\sqrt{s}$, where $s$ is the number of pixels
used in the smoothing kernel. In this case, $N = 52$, $n = 5$, and $s
= 9$, so $(Ns/n)^{1/2} \approx 10$.

\section{Discussion}
\label{sec:discussion}

\begin{figure}
\centering
\begin{tabular}{cc}
\includegraphics[scale=0.75]{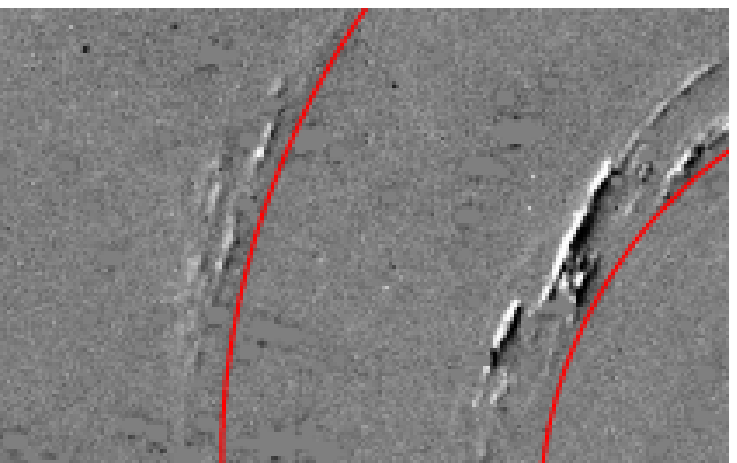}
  & \includegraphics[scale=0.75]{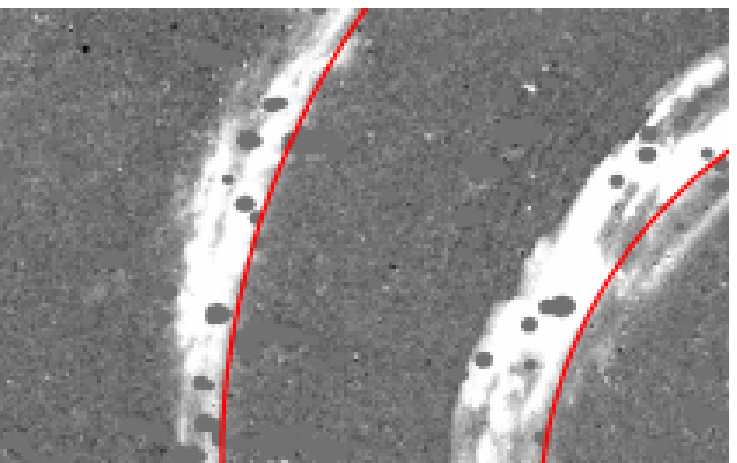} \\
\includegraphics[scale=0.75]{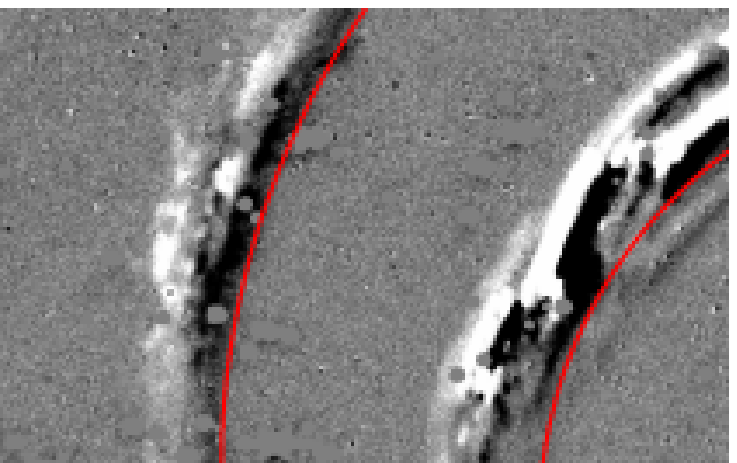}
  & \includegraphics[scale=0.75]{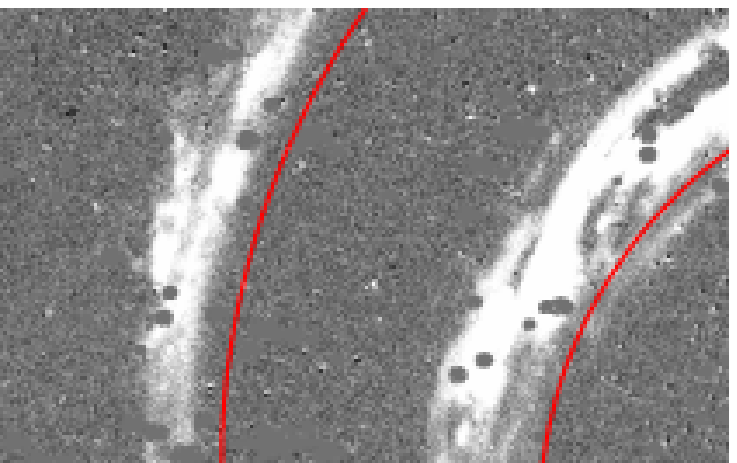} \\
\includegraphics[scale=0.75]{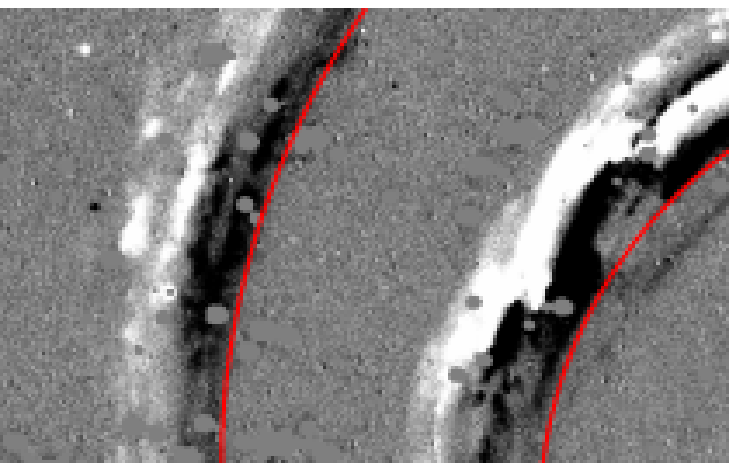}
  & \includegraphics[scale=0.75]{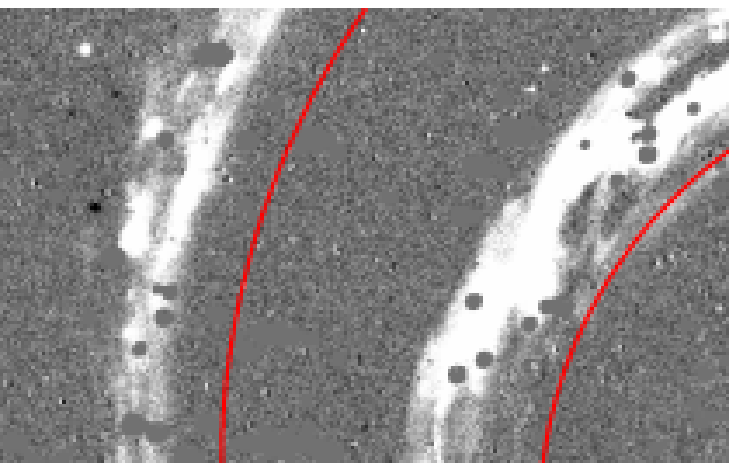} \\
\includegraphics[scale=0.75]{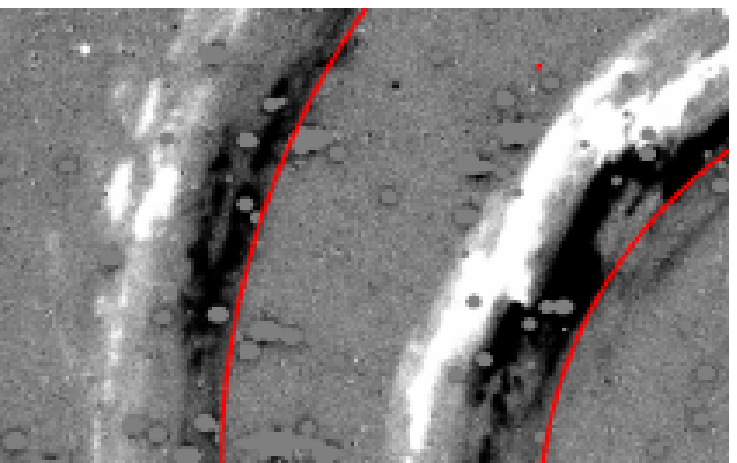}
  & \includegraphics[scale=0.75]{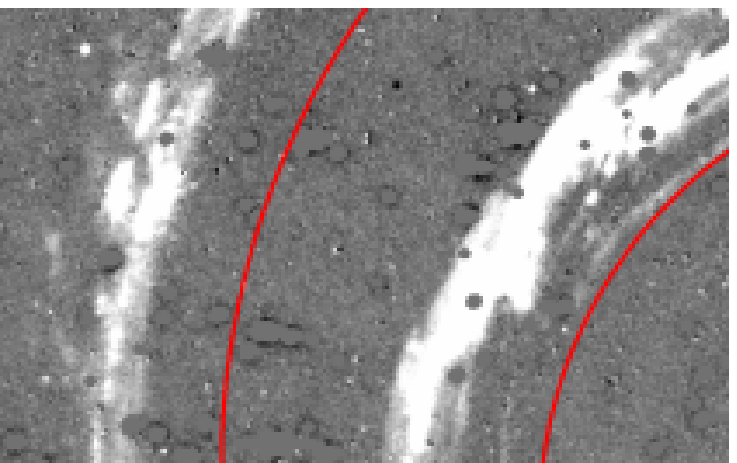} \\
\includegraphics[scale=0.75]{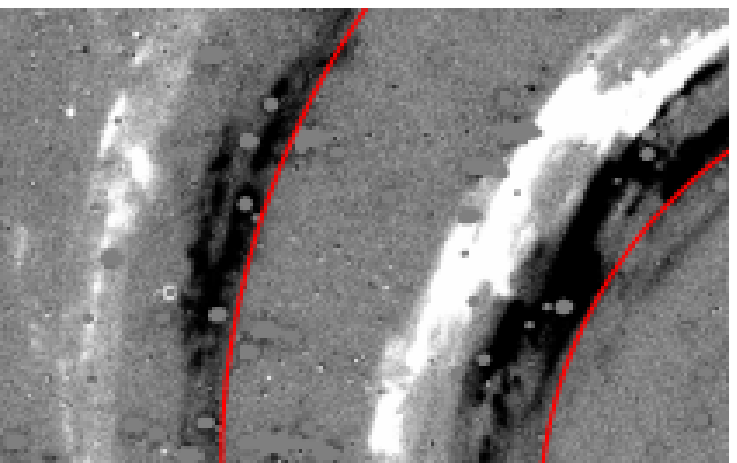}
  & \includegraphics[scale=0.75]{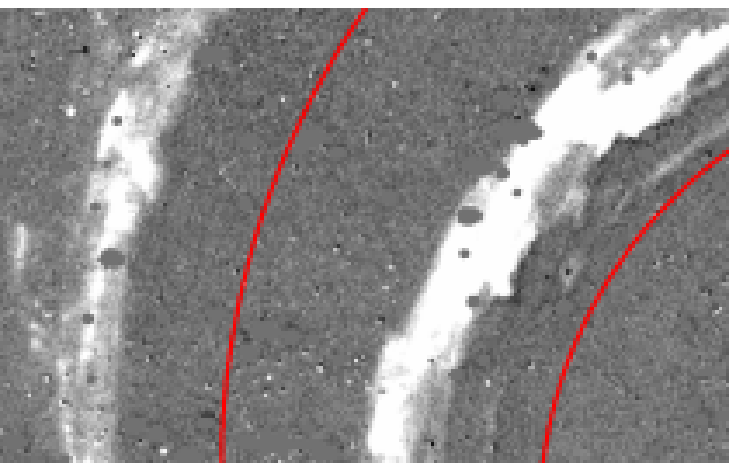}
\end{tabular}
\caption{
\emph{Left column:} Difference images with template epoch 2001 November 15 and image epochs
2001 December 11, 2002 December 14, 2003 December 15, 2004 December 15,
and 2005 December 21, respectively,
from top to bottom. \emph{Right column:} Zero-flux--corrected images for the
same five image epochs. The red arcs provide a constant point of reference.
In each difference image, the light echo positions
in the image and template epochs overlap, causing partial or complete cancellation.
The NN2 algorithm disentangles this information to reveal the complete extent of the light
echoes. Note especially in the first image the near complete cancellation evident
in the difference image, in which the image and template dates are separated by
only 1 month. Compare this to the image on the right. The FOV is
$3.0^{\prime} \times 2.0^{\prime}$.\label{fig:lightechotimeseries}}
\end{figure}

Figure~\ref{fig:lightechotimeseries} shows the evolution of one light
echo region over our five seasons of observation. The left column shows difference
images with the same template epoch in 2001 November.  The image epochs are
separated by 1 year and fall in December 2001, 2002, 2003, 2004 and
2005. The right column shows the zero-flux--corrected NN2 images for the
same five epochs. Note that in the first difference image, the
template and image dates are separated by less than 1 month. The
light echoes move very little during this short interval, so the
difference image shows almost complete cancellation between the light
echoes at the two epochs. However, the accompanying zero-flux--corrected NN2
image shows the full extent of the light echo. The second
and subsequent difference images show some regions that appear to
contain flux only from the image epoch, and others that appear to
contain flux only from the template epoch. However, it is difficult
to make these judgments with certainty, and there are always
significant regions of overlap, in which the difference image reveals
only the change in flux. The NN2 algorithm allows us to disentangle
this information and extract the relative light echo fluxes across a
range of epochs rather than only two. It is important to note that
the images in the right column of Figure~\ref{fig:lightechotimeseries}
are not obtained directly from the accompanying difference images, but
rather depend on all $N(N-1)/2$ difference images obtained from the
$N$ observations. This figure clearly shows the limitations of the
single-template difference imaging method and the advantages offered
by the NN2 algorithm.

In addition to uncovering the spatial extent of light echoes, our technique
can reveal faint light echo features not visible in the difference
images. Figure~\ref{fig:weakringseries} shows stages of analysis for a
region containing a faint light echo ring in the left part of the
frame. Note that this ring is barely tracable in the
sample difference image. In our full mean-combined images, light echo
features with peak surface brightnesses of 25 mag arcsec${}^{-2}$
are clearly visible, whereas the light echo detection limit in the
difference images is about 24 mag arcsec${}^{-2}$.

Our code, along with images from an example region around SN 1987A,
are available online.\footnote{See \texttt{\url{http://www.ctio.noao.edu/supermacho/NN2}}.}
The NN2 calculations involve several large matrix inversions and are CPU-intensive.
Our Python code required roughly 30 hours on 20 CPUs to process 52 images of
dimensions $4160 \times 1100$. Although vast speed improvements can likely be
expected from C code, the processing demands should be borne in mind when
considering larger scale applications. 

\section{Conclusions}
\label{sec:conclusions}
Difference imaging is a standard tool in the analysis of light
echoes. Without a light echo-free template, however, difference images
contain overlapping light echo fluxes from two epochs. We present here a new method that
produces images containing absolute light echo fluxes from one epoch
only. The NN2 algorithm is used to calculate a relative light curve
across a range of $N$ epochs by considering difference images
constructed from every pair of epochs. By applying this method over a
region, we obtain light curves for each pixel. However, these lightcurves
have arbitrary offsets with respect to each other and to the
true zero-flux level. Hence, we apply a statistical method to estimate
the true zero-flux level for each pixel and shift the light curves
accordingly. This method is optimized for detecting faint light echoes
with peak surface brightnesses as faint as 25 mag arcsec${}^{-2}$.
By spatially and temporally binning the images, we greatly increase
the signal-to-noise ratio. Our technique is capable of revealing fine
structure and faint light echo regions. It requires no privileged or
light echo-free images and can be applied completely automatically to
a wide variety of light echo situations. In addition, variations on this
technique could permit analysis of other extended variable light sources,
such as stellar outflows and supernova remnants.
Future work will include applications to recently discovered light echoes from
ancient supernovae in the Large Magellanic Cloud \citep{Rest05:le}.

\section{Acknowledgments}

The SuperMACHO survey is being undertaken under the auspices of the
NOAO Survey Program. We are very grateful for the support provided to
the Survey program from the NOAO and the National Science
Foundation. A.~Rest thanks NOAO for the Goldberg
Fellowship. SuperMACHO is supported by the STScI grant GO-10583 and
GO-10903.  Discussions with Michael Wood-Vasey were very
valuable. A.~Newman thanks the National Science Foundation for their
support via the Research Experience for Undergraduates (REU) program.
We thank the referee B.~Sugerman for very useful comments and suggestions
which helped improve the quality of this paper.

\bibliographystyle{apj}
\bibliography{ms}
\clearpage

\clearpage

\clearpage

\clearpage

\clearpage

\clearpage

\clearpage

\clearpage

\end{document}